%% file: Thermoelectrics.tex
\documentclass[graybox]{svmult}

\usepackage{mathptmx}       % selects Times Roman as basic font
\usepackage{helvet}         % selects Helvetica as sans-serif font
\usepackage{courier}        % selects Courier as typewriter font
\usepackage{type1cm}        % activate if the above 3 fonts are
\usepackage{makeidx}         % allows index generation
\usepackage{graphicx}        % standard LaTeX graphics tool
\usepackage{multicol}        % used for the two-column index
\usepackage[bottom]{footmisc}% places footnotes at page bottom

\makeindex             % used for the subject index
\newcommand{\vk}{{\mathbf{k}}}

\begin{document}

%\title*{Emergent Mottness and Thermoelectricity}
%Modern title. Conference title.
\title*{Thermoelectrics Near the Mott Localization-Delocalization
Transition}
\author{Kristjan Haule and Gabriel Kotliar}
\institute{Kristjan Haule \at Physics Department and Center for Materials Theory,
  Rutgers University, 136 Frelinghuysen Road, Piscataway, NJ, \email{haule@physics.rutgers.edu}
\and Gabriel Kotliar \at Physics Department and Center for Materials Theory,
  Rutgers University, 136 Frelinghuysen Road, Piscataway, NJ, \email{kotliar@physics.rutgers.edu}}
\maketitle

\section{Current status of Material Design Using Correlated Electron Systems}
\label{sec:1}

\subsection{Weakly Correlated Systems}

The dream of accelerating the discovery of materials with useful
properties using computation and theory is quite old, but actual
implementations of this idea are recent \cite{Picket,Hafner}.
Successes in material design using weakly correlated materials, are
due, to a large degree, to a two important developments:
\begin{itemize}
\item[a)] approximate implementations of first
principle methods, which are relatively accurate and
computationally efficient
\item[b)] robust implementation of algorithms
which are highly reproducible and widely available in well tested
codes.
\end{itemize}
Density functional theory based approaches give reliable estimates of
the total energy, and are an excellent starting point for computing
excited state properties of weakly correlated electron systems.  These
approaches allows the evaluation of transport coefficients using very
limited, or no empirical information, and are beginning to be used in
conjunction with data mining techniques and combinatorial searches.

\subsection{Strongly Correlated Electron Systems}

Since a large number of interesting physical phenomena, such as high
temperature superconductivity and large Seebeck coefficients, are
realized in strongly correlated electron systems, there is a great interest
in the possibility of carrying out rational material design with
correlated materials \cite{Picket,Hafner}.

The theoretical situation in this area, however, is a lot more
uncertain. For example, the issue of whether the two dimensional
one band Hubbard model supports superconductivity or not is still
very open \cite{imada}. Given that this model is an extraordinary
oversimplification of realistic materials, it is hard to
contemplate explaining, let alone predicting experimental results
in materials that require a much more elaborate models for their
description.  The prospect of predicting properties of materials
which have not yet been synthesized is even more daunting. In
this chapter we will argue that this assessment is overly
pessimistic, and we will give some reasons why we expect a rapid
progress in the coming years through the interplay of qualitative
reasoning, new theoretical and computational methods, and
experiments. We will then describe some attempts to gain
experience in this field, and some lessons that we have learned
using thermoelectric performance as an example.

\subsubsection{Dynamical Mean Field Theory}

The advent of  Dynamical Mean Field Theory (DMFT) removed many
difficulties  of the traditional  electronic structure methods.
DMFT  describes Mott insulators, as well as correlated metals.
DMFT combines ideas of quantum chemistry, such  performing a full
configuration interaction calculation (at a local level to avoid
size consistency problems),  and physics, such as  carrying out a
diagrammatic expansion  around the band limit. DMFT treats
quasiparticle bands and Hubbard bands on the same footing, and,
unlike simpler approaches such as LDA+U, is able to describe the
multiplet structure of correlated solids. The latter is being
inherited from open shell atoms and ions. DMFT has been
successful in accounting for the behavior observed in correlated
materials ranging from plutonium to vanadium oxides and have even
made some predictions, which have been successfully confirmed by
experiment\cite{rmp1,vollhardt}.  This suggests that the approach
is reasonably accurate, in the sense that it gives a zeroth order
picture of correlated materials  not too close to criticality. Ten
years ago, a combination of DMFT with electronic structure
methods  LDA+DMFT, was proposed \cite{LDA+DMFT1,rmp2,held} and
accurate implementations are being actively developed across the
world. Just like LDA, these tools connect the atomic positions
with the physical observables using very little information from
experiment, and therefore they have the potential to accelerate
material discovery.

Predicting the phase diagram of strongly correlated materials is
an extremely difficult problem. Correlated materials have many
competing phases, which are very close in energy.  This poses
serious difficulties to direct numerical studies of model
Hamiltonians  because omitting small terms  which are present in
the Hamiltonian of  the actual material  or finite size effects
connected to boundary conditions  can exchange the stability of
two very different phases.

DMFT divides the solution of the many body problem of a solid
state system into two separate and distinct steps: the study of
the evolution of the mean field solutions as a function of
parameters and the computation of total energies for each  DMFT
solution. Common to many mean field approaches, a given
Hamiltonian can have many distinct DMFT solutions, describing
various possible phases of a material. Which phase is realized
for a given value of parameters (temperature, volume, stress,
doping concentration of impurities, etc.) is determined by
comparing the free energy of the different DMFT solutions. A lot
of important information can be obtained from the first step
alone, when combined with experimental information. If one knows
that for some value of parameters certain phase is realized in
material, one can use DMFT to explore the properties of that
phase, and optimize desired physical property, sidestepping the
difficult issue of the comparing the free energies of the
different  competing phases which can be done  at a later stage.
The rational  material design process  should then suggest
modifications of the  material to stabilize the phase with
desirable properties.

\begin{figure}[b]
\sidecaption
\includegraphics[scale=0.5]{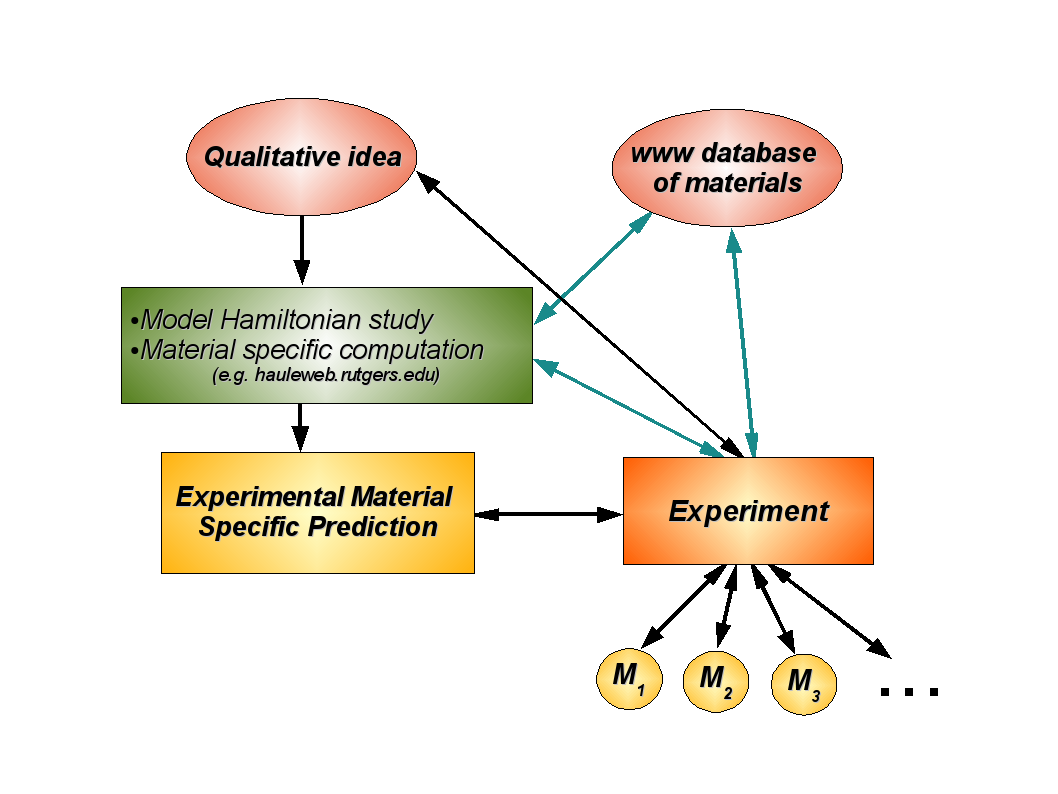}
\caption{ A schematic drawing of the rational material design
process. It relies on condensed matter theory, material databases
and realistic DMFT implementations,  and   it involves a close
and iterative interplay of theory and experiment. } \label{fig:1}
\end{figure}

\section{The process of rational material design}

Figure~\ref{fig:1} describes schematically the rational material
design process. It begins with a qualitative idea, which is then
tested by a calculation. One of the major advances of realistic
DMFT implementations such as LDA+DMFT or GW+DMFT is that now this
calculation can be made material specific, resulting in a set of
predictions  that can be tested experimentally . The experimental
results can either rule out the qualitative idea, in which case
the process stops, or reinforce and refine the idea. Experiments
also help calibrate the computational methods, which in turn lead
to an improved material specific prediction in the next iteration.
The expected result form this process are materials with improved
properties $M_1$, $M_2$ , $M_3$, $\cdot$. In addition, this
process tests theoretical ideas in an unbiased way, deepens our
understanding of materials physics, and refines the accuracy of
computational tools. Large databases of existing materials are
being  created (for example http://icsd.ill.eu/icsd/index.html),
which are starting to be used, in combination with the first
principle methods, for data mining techniques \cite{ceder}. Using
the crystal structure information from the database. the first
principles methods can identify potentially promising materials,
which can then be analyzed experimentally.

\section{Thermoelectricity of Correlated Materials}

\subsection{Formalism}

The transport coefficients that govern the thermopower, electric and
thermal conductivity can be expressed in terms of the matrix of
kinetic coefficients $A_{m}$ relating the electric and thermal
currents $J$, $J_{Q}$ to the applied external fields $\nabla \mu /T$,
$\nabla T/T^{2}$.  Transport quantities become
$S=-(k_B/e)(A_{1}/A_{0})$, $\sigma =(e^2/T) A_{0}$, $\kappa =
k_B^2[A_{2}-{{A_{1}}^{2}/A_{0}}]$.  The thermoelectric response thus
reduces to the evaluation of kinetic coefficients.

The thermoelectric figure of merit is defined by
\begin{equation}
ZT=\frac{S^{2}\sigma T}{\kappa+\kappa_{phonon} } \label{ZT0}
\end{equation}
where $T$ is the absolute temperature, $\sigma $ is the electrical
conductivity, $S$ is the Seebeck coefficient or thermopower, and
$\kappa$ ($\kappa_{phonon}$) is the electron (phonon) contribution to
the thermal conductivity.

The Wiedemann Franz law is an approximate relation that allows us to
estimate the ratio of the electronic contribution to the thermal
conductivity ($\kappa$) and electric conductivity ($\sigma$). It
postulate that the Lorentz number, $L=\kappa/(\sigma T)$, is weakly
material dependent.

Its value at low temperatures is given by $(\pi^2/3) (k_B/e)^2 =
2.44\times 10^{-8} W\Omega/K^2$.  We will return to the Lorentz
number at higher temperatures later in this article.  If we
ignore the thermal conductivity of the lattice, the figure of
merit can be written as $ZT = S^2 /L$, hence to have a promising
figure of merit ($ZT$ close to or larger than one) it is
necessary to have $S$ bigger than the basic scale $k/e = 86 10^6
V/K$.  The thermal current of an interacting electronic system
was determined first by Mahan and Jonson \cite{mahan}.  Ref
\cite{mahan} discusses a model containing electrons interacting
with phonons, and the review \cite{mahan_review}  discusses the
general case of the electron electron interactions (see also
ref~\cite{paul}).

DMFT expresses the one particle Greens function in terms of a
local self energy of an impurity model, satisfying a self
consistency condition. Practical evaluation of the transport
coefficients becomes possible in the approximation on small
vertex corrections. This was first done by H. Schweitzer and G.
Czycholl \cite{czycholl} (see also ref \cite{pruschke}).  For the
Hubbard like interactions, there are no contributions from the
non local Coulomb interactions, and the negligence of the vertex
corrections can be justified rigorously in the limit of infinite
dimensions \cite{khurana}.  The same is true, but far less
obvious, for the thermal current, as it was shown in
ref~\cite{paul}.  In the multiorbital situation, the vertex
corrections to the conductivity need to be examined on a case by
case basis, and do not necessarily vanish, even in infinite
dimensions. With this approximation, the LDA+DMFT transport
coefficients reduce to
\begin{eqnarray}
\label{eq:A_coefficients2}
A_{m}^{\mu\nu} = \pi T \int{d\omega}\left(-\frac{df}{d\omega}\right)\left(
\frac{\omega}{T}\right)^m \sum_\vk\mbox{Tr}[v_\vk^\mu(\omega)\rho_\vk(
\omega)v_\vk^\nu(\omega)\rho_\vk(\omega)]
\end{eqnarray}
where $\textbf{v}_{\vk\; ij}=-\frac{i e}{m}\langle\psi_{\vk i}|\nabla|\psi_{\vk j}\rangle$ are velocities of electrons and $\rho_\vk$ is the
electron spectral density
\begin{equation}
\rho_\vk(\omega)=\frac{1}{2\pi i}[G^\dagger_\vk(\omega)-G_\vk(\omega)].
\end{equation}

The weakly interating case appears as a limiting case where the
spectral function becomes a delta function $\rho_{\vk\; ij}(\epsilon) =
 \delta(\epsilon - \epsilon_{\vk i})\delta_{ij}$.  One can therefore
formulate the problem of the optimization of the figure of merit as
the problem of optimizing a functional of spectral functions, with
self energies which are realizable from an Anderson impurity model,
with a bath satisfying the DMFT self condition.

\subsection{Thermoelecricity near the Mott transition: qualitative  considerations}

Following the early developments of DMFT and its successful
application to the theory of the Mott transition in three
dimensional transition metal oxides (for reviews see \cite{rmp1}
\cite{vollhardt}), it was natural to use this approach to
formulate and answer the question of whether we should look for
good thermoelectrics near the Mott localization delocalization
transition.  This talk describes our current understanding of
this  issue and the tentative answer, at this point,  is no, but
perhaps yes.

\begin{figure}[b]
\sidecaption
\includegraphics[scale=0.6]{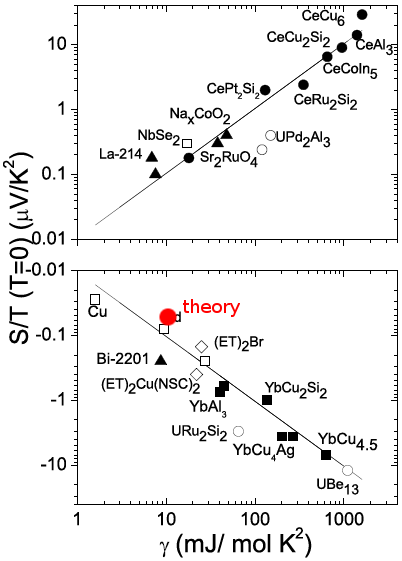}
\caption{ Bhenia Jaccard Flouquet plot from \cite{behnia}
\label{behnia}. The theoretical point obtained on the LaSrTiO3
system with twenty percent doping away from the Mott insulator is
also shown in the same graph. } \label{fig:2}
\end{figure}

There were several reasons to suspect that proximity to the
localization delocalization transition is good for thermoelectricity:
\begin{itemize}
\item[a)] Sharp structures in the density of states lead to large $S$
  in simple theorites \cite{sofo}. The modern theory of the Mott
  transition predics a quasiparticle peak, which narrows as the
  transition is approached. And this could result a large
  thermoelectric response.
\item[b)] One can think on a qualitative level of the thermoelectric
  coefficient as the entropy per carrier. In the incoherent regime,
  one could imagine that each carrier can transport a large amount of
  entropy.  The incoherent regime, above a characteristic coherence
  temperature $T^*$, is easy to access near a localization
  delcolization transition, because the proximity tot this boundary
  makes $T^*$ low.
\item[c)] Orbital degeneracy increases the number of carriers and
  would be expected to increase the figure of merit. There are many
  orbitally degenerate three dimensional correlated transition metal
  oxide.
\end{itemize}

Ref.~\cite{gunnar} considered a model of the prototypical doped
insulator LaSrTiO$_3$, which has been carefully investigated
experimentally  \cite{tokura}. The thermoelectric properties of
this systems had not been investigated at that time. Early DMFT
studies accounted for the divergence of the linear term of the
specific heat, and the susceptibility, as well as the existence
of a quasiparticle peak in the spectra \cite{Fujimori}.

The Hall coefficient, however, coincides with the band theory
calculations, and is non critical near the Mott transition
\cite{kajuter}.  It is possible to analyze  the DMFT transport
equations in two regimes: i) $T \ll T^*$, where the electronic
transport is controlled by band-like coherent quasiparticles,
well described in momentum space, ii) $ T \gg T^*$ when the
electron is better described as a particle in real space, and the
transport is diffusive \cite{gunnar} (see below).

The second regime is well described by the high temperature
expansion, valid for $T > D$ ($D$ is the bandwidth). Approximate
numerical solutions of the DMFT equations Ref.~\cite{gunnar}
showed  that the thermoelectric response computed in  the high
temperature regime could be  matched smoothly with the low
temperatures results  valid for $T \ll T^*$.

\subsection{Application to LaSrTiO$_3$}

\begin{figure}[b]
\sidecaption
\includegraphics[scale=0.5]{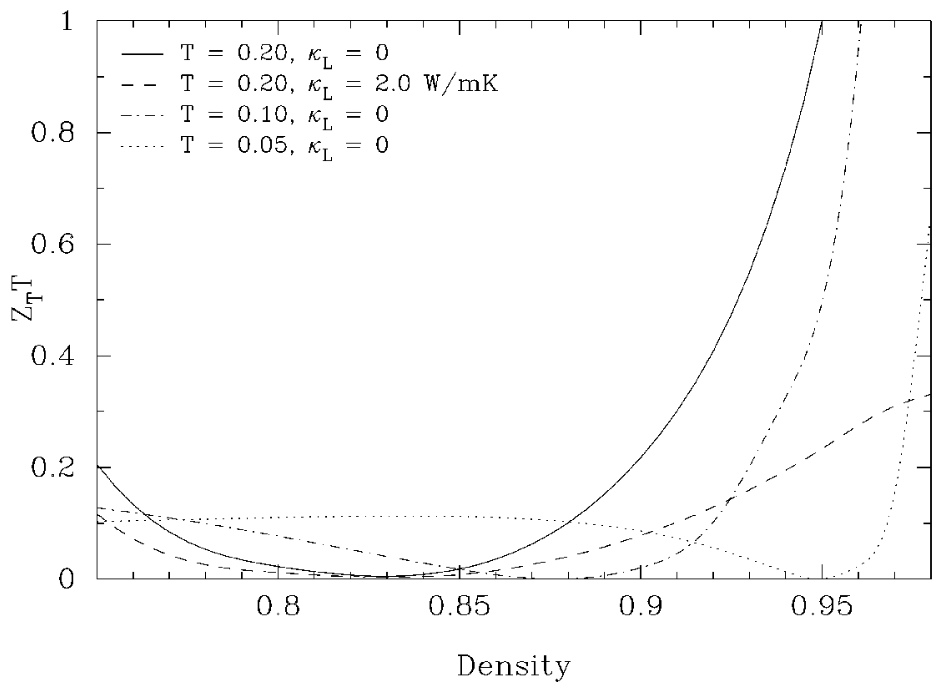}
\caption{ Figure of merit  for different values of the lattice
thermal conductivity. The electron density is $1-x$.}
\label{fig:5}
\end{figure}

\begin{figure}[b]
\sidecaption
\includegraphics[scale=0.42]{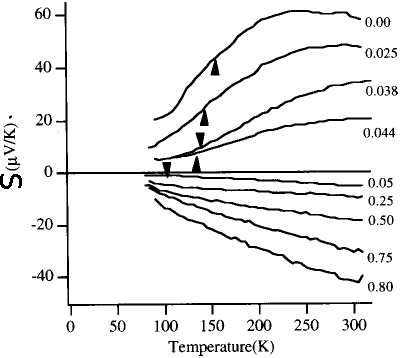}
\includegraphics[scale=0.42]{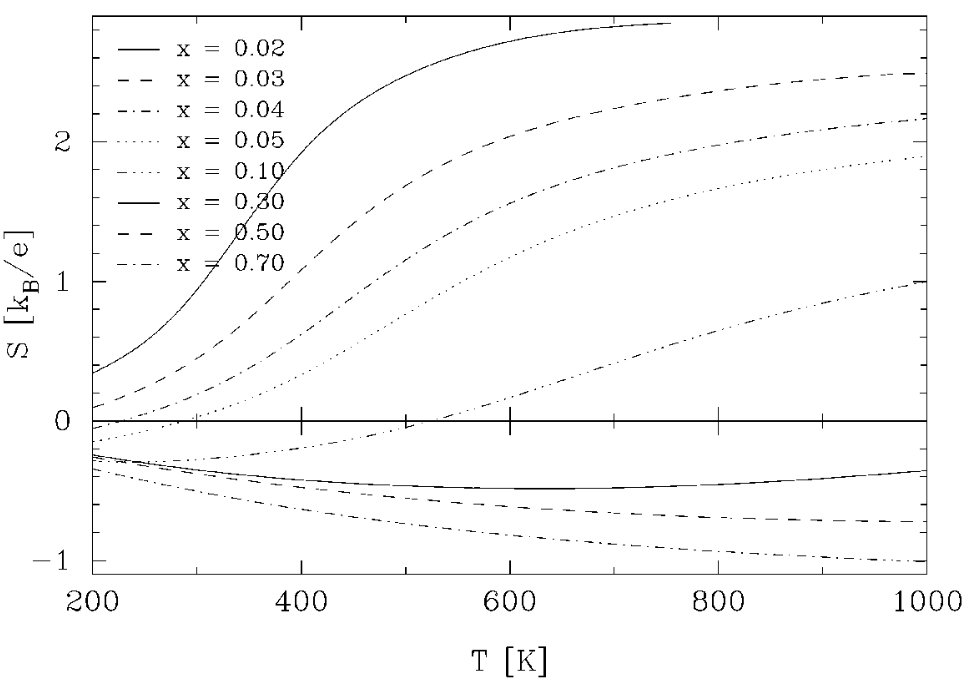}
\caption{ Experimental (left panel) and theoretical computations
of the thermoelectric power ($S$) of the La$_{1-x}$Sr$_x$TiO$_3$ from
Refs.~\protect\cite{{gunnar}} and \protect{\cite{goodenough}}.
}
\label{fig:3}
\end{figure}

An approximate numerical solution of the DMFT equations for the
titanides was shown to interpolate smoothly between the high
temperature and low temperature region. This is consistent with
the idea that DMFT reconciles the band picture at low energies
and low temperatures, with the particle picture at high energies
and high temperatures.  The temperature scale here is set by the
coherence temperature $T^*$. Taking a tight binding
parametrization suitable for the titanites, the thermoelectric
figure of merit as a function of temperature and doping is
reproduced in Fig.~\ref{fig:5}.

The behavior of the thermoelectric power near the Mott transition
is shown in Fig.~\ref{fig:3}. Notice that at low doping, the
contribution from the lower Hubbard band dominates and the
thermoelectric power is positive while at high doping the
quasiparticle contribution dominates and the thermoelectric power
is electron-like. Measurements near the Mott transition were
carried out a few years later \cite{goodenough}, and they are
qualitatively, but not quantitatively, similar to the theory.
This is to be expected, given the various approximations that
were made (the electronic structure, the lattice distortion, and
crystal field effects ignored, the impurity solvers used were very
approximate). More recent  studies of these materials including
lattice distortions and crystal field splittings  have been
carried out\cite{pavarini} but their effects on the thermoelectric
response has not been studies.

\subsection{Low Temperature Regime}

LaSrTiO$_3$ is described by a multi-band Hubbard model. At low
temperatures, the Fermi liquid theory is valid. The  slope of the
real part of the self energy scales as $1-1/Z$, where $Z$ is the
quasiparticle residue. The quadratic part of the self energy
is related to the quasiparticle lifetime, which is small in the
Fermi liquid regime.

Under these assumptions, we can rewrite a simpler expressions for the
transport coefficients $A_n$ of a multiband Hubbard model at low
temperatures
\begin{equation}
A^{\mu\nu}_n = {\frac{N k_B T}{8}}
  \int_{-\infty}^{\infty} dx \frac{x^n}{\cosh^2(x/2)}
      {\frac{\Phi_{\mu\nu}(x T +\mu -\Sigma'(x T))}{\Sigma''(xT)}},
\label{simpler}
\end{equation}
where
$\Phi_{\mu\nu}$ is the transport function defined by
$\Phi_{\mu\nu}=\sum_\vk v_\vk^\mu
v_\vk^\nu\delta(\omega-\epsilon_\vk)$ and $\Sigma''(\omega)$ is the
imaginary part of the electron self-energy.

At low temperatures, $A_0$ and $A_2$ are simply estimated
by replacing $\Sigma''(\omega)$ by its quadratic approximation,
$\Sigma''(\omega) \sim \frac{\gamma_0}{Z^2} (\omega^2 + \pi^2
T^2)\equiv\Sigma^{(2)}(\omega)$. We then obtain
$$ A_{2n} = \frac{Z^2}{T}\frac{N k_B}{2\gamma_0\pi^2}\; E_{2n}^1 \Phi_{\mu\nu}(\mu_0)$$
where $\mu_0=\mu-\Sigma'(0)$ and
$$E_n^k = \int_{-\infty}^{\infty} \frac{x^n dx}{4\cosh^2(x/2)[1+(x/\pi)^2]^k}$$
are numerical constants of the order unity.

On the other hand, this approximation neglects particle-hole asymmetry
and gives zero thermoelectricity since $E_1^1=0$. There are two
sources of particle hole asymmetry. One is obtained by
expanding the transport function in Eq.~\ref{simpler} to the first
order, which describes the particle hole asymmetry in the electronic
velocities, contained in the bare band structure of the problem.
This term can be approximated by the LDA Seebeck coefficient divided
by quasiparticle renormalization amplitude $Z$. The second
contribution is the result of the particle hole asymmetry of the
scattering rate. It involves subleading {\it cubic} terms in the self
energy, which scale near the Mott transition as
\begin{eqnarray}
&&\Sigma''(\omega)= \Sigma^{(2)}(\omega) +\Sigma^{(3)}(\omega)\nonumber+\cdots\\
&&\Sigma^{3}(\omega)= \frac{(a_1 \omega^3 +a_2\; \omega\, T^2)}{Z^3}
\label{correction}
\end{eqnarray}
and $a_1$, $a_2$ are constants of order unity (even terms in frequency are not important).
This leads to the following expression for the thermoelectric coefficient
\begin{equation}
A_1 = Z \frac{ N k_B }{2\gamma_0\pi^2}
\left[\Phi'_{\mu\nu}(\mu_0)\; E_{2}^1
  -\Phi_{\mu\nu}(\mu_0)\; (a_1 E_4^2 + a_2 E_2^2)/\gamma_0
  \right]
\end{equation}
where $\Phi'(x)=d\Phi(x)/dx$.

Unfortunately it has proved to be very difficult to estimate the
magnitude of the coefficients $a_1 $ and $a_2$. It is important to
develop intuition into when these terms are important and their
sign. Since in many cases, LDA predicts the correct sign of the
thermoelectric power at low temperatures, perhaps the scattering time
particle hole asymmetry Eq.~(\ref{correction}) is not dominant in the LaTiO$_3$
system. This problem deserves a thorough investigation.

At low temperature, the thermoelectric
coefficients is
\begin{equation}
S = -\frac{k_B}{|e|}\frac{k_B T}{Z}
\left[\frac{\Phi'(\mu_0)}{\Phi(\mu_0)}\frac{E_2^1}{E_0^1}-
  \frac{a_1 E_4^2+a_2 E_2^2}{\gamma_0 E_0^1}\right],
\end{equation}
which clearly scales as $T/Z$ with $Z$ vanishing at the Mott
transition. Since the linear term of the specific heat $\gamma$ scales
as $1/Z$ the ratio $S/(\gamma T)$ in a Hubbard like
model approaches a finite value as $Z$ vanishes:
\begin{equation}
\frac{S}{\gamma T} = -\frac{3}{|e|}\frac{1}{D(\mu_0)}
\left[\frac{\Phi'(\mu_0)}{\Phi(\mu_0)}
  \frac{E_2^1}{E_0^1}-\frac{a_1 E_4^2+a_2 E_2^2}{\gamma_0 E_0^1}\right].
\end{equation}
The first part of the ratio depends only on the bare band-structure
quatities and is not effected by strong correlations. The second part,
however, is due to the asymmetry of the quasiparticle lifetime, and
might be less universal and more material and correlation
specific. This question deserves further study.

For the LaSrTiO$_3$ system, we estimated its value numerically using
LDA+DMFT \cite{udo} and we include its value in the plot of Behnia
et.al. \cite{behnia} in Fig.~\ref{fig:2}. In Ref.~\cite{behnia} it was
observed that the ratio $ S / {\gamma T} $ is weakly material
dependent in a large number of materials which they compiled.
From the theoretical point of view, the weak dependence of the ratio
of Behnia et. al. on material can be view as a validation of the local
approximation, since the most material dependence is embodied in the
quantity $Z$, which cancels in the ratio $S/(\gamma T)$.
This suggest that the DMFT approach holds great promise
for the search of good thermoelectric materials.  Deviations from
universality arise from the variations of the bare density of states
and from the effects of the cubic terms in the self energy that were
not included in the analysis of ref \cite{gunnar}.  It would be
interesting to return to this problem using modern LDA+DMFT tools.

\subsection{High temperature results}

In the high temperature region, the expansion of the solution of the
DMFT equations led to the generalized Heikes formula
\cite{Chaikin,Koshibae} for the Seebeck coefficient. In this limit,
thermopower is given by $S = \mu/(e T)$, where $\mu$ is the chemical
potential.  The exact diagonalization of the atomic problem gives a
set of atomic eigenvalues $E_m$ and their degeneracies $d_m$. The
chemical potential is then determined from the partition sum
\begin{equation}
n = \sum_{m} d_m e^{-\beta (E_m - \mu N)}/Z
\end{equation}
where $n$ is the number of electrons in a correlated orbital.
Hence, valence of the solid $n$ can be used to predict the high
temperature value of thermopower.

For the case of $n\le 1$, which is relevant for the titanides, the
expressions for transport quantities take the explicit form:
\begin{eqnarray}
\sigma &=& \frac{e^2}{a\hbar}\pi N(D\beta)\gamma_0\frac{\frac{n}{N}(1-n)}{[\frac{n}{N}+(1-n^2)]^2}\\
S &=& \frac{k_B}{e}\log\frac{n}{N(1-n)}\\
\kappa &=& \frac{k_B D}{a\hbar}\pi N(D\beta)^2 \gamma_2 \frac{\frac{n}{N}(1-n)}{[\frac{n}{N}+(1-n)]^2}.
\end{eqnarray}
Here $N$ is the spin and orbital degeneracy, and $n$ is the electron
density, $D$ is half of the bare bandwidth and $\gamma_0$, $\gamma_2$
are numerical constants of order unity.

Notice that at high temperature the Lorentz number is given by $L=
(k/e)^2 (D/kT)^2 \gamma_2 / \gamma_0$ with $\gamma_2 $. Hence the
Lorentz number in a model with a fixed number of particles and finite
bandwith goes to zero at high temperatures. Thus eventually the the
electronic thermal conductivity becomes less than the lattice
conductivity and the latter controls the figure or merit.  This effect
was modeled in the dashed curve of Fig.~\ref{fig:5}, where the effects
of the lattice thermal conductivity was modeled by a constant
$2.0$W/mK. The inclusion of the lattice thermal conductivity resulted
in a dramatic reduction of the figure of merit.
We can interpret the high temperature DMFT results for the thermal
transport using a well known equation $\kappa= \frac{1}{3} v_F c_V l$,
where $v_F$ is the Fermi velocity, $c_V$ the specific heat, and $l$
the electron mean free path.
Since the specific heat decreases as $(D/T)^2$, the mean free path
has saturated to a lattice spacing, and the velocity of the electrons
is of the order of $v_F$. This is consistent with the value of the
conductivity if one uses the Einstein relation
$\sigma=D_c\,{ dn/{d \mu}}$
with $ {dn/ {d \mu}} \approx 1 /T $ and the charge diffusion
constant $D_c $= ${v_F} l$. Here  the mean free path $l$ is of the order of
the lattice spacing, and the Fermi velocity $v_F$ is approximately
temperature independent.

\section{Towards Material Design}

\subsection{Rules for good correlated thermoelectricity}

From the theoretical analysis it becomes clear why
La$_{1-x}$Sr$_{x}$TiO$_3$ is not a good thermoelectric material near
the Mott transition. The contributions from the Hubbard bands and the
quasiparticle peak have opposite signs, and they compete with each
other in the interesting temperature regime, when $T$ is comparable to
$T^*$. This observation leads to empirical rules for the search for
good correlated thermoelectric materials:

\begin{itemize}
\item[(1)] The optimal performance (when the thermal conductivity
  of the lattice is taken into account) occurs in the crossover region
  $ T \approx T^*$.  Hence one should tune $T^* $ to the temperature
  region where the thermoelectric device operates.  One should also
  reduce the electronic thermal conductivity (and therefore also the
  electric conductivity) until it becomes comparable to the lattice
  thermal conductivity, but not any further.

\item[(2)] In the crossover regime, both the quasiparticle bands and
  the Hubbard bands contribute to the transport. Hence one should try
  to optimize {\it both} high temperature and low temperature
  expressions for the figure of merit. Therfore good candidates for
  thermoelectricity should have quasiparticle carriers and Hubbard band
  carriers of the same sign.
\end{itemize}

We see that LaSrTiO$_3$ does \textit{not} satisfy the second rule,
and hence its figure of merit is not large.  The quasiparticle
contribution to the thermopower is electron-like while the lower
Hubbard band contribution is hole like.

It is instructive to contrast the titanites with the cobaltate
materials  which have a larger thermoelectric response.  The
cobaltates have holes in the lower Hubbard band while the
quasiparticle contribution evaluated from the LDA \cite{singh}
has a positive sign, hence it satisfies the second rule for good
thermoelectricity.

An investigation of the density driven Mott transition in the
context of a two band Hubbard model, with one electron per site,
was carried out in Ref.~\cite{udo2}, and the conclusions are  very
similar to those obtained in  the context of the  doping driven
Mott transition.

\subsection{Emergent Mottness}

\begin{figure}[b]
\sidecaption
\includegraphics[scale=0.42]{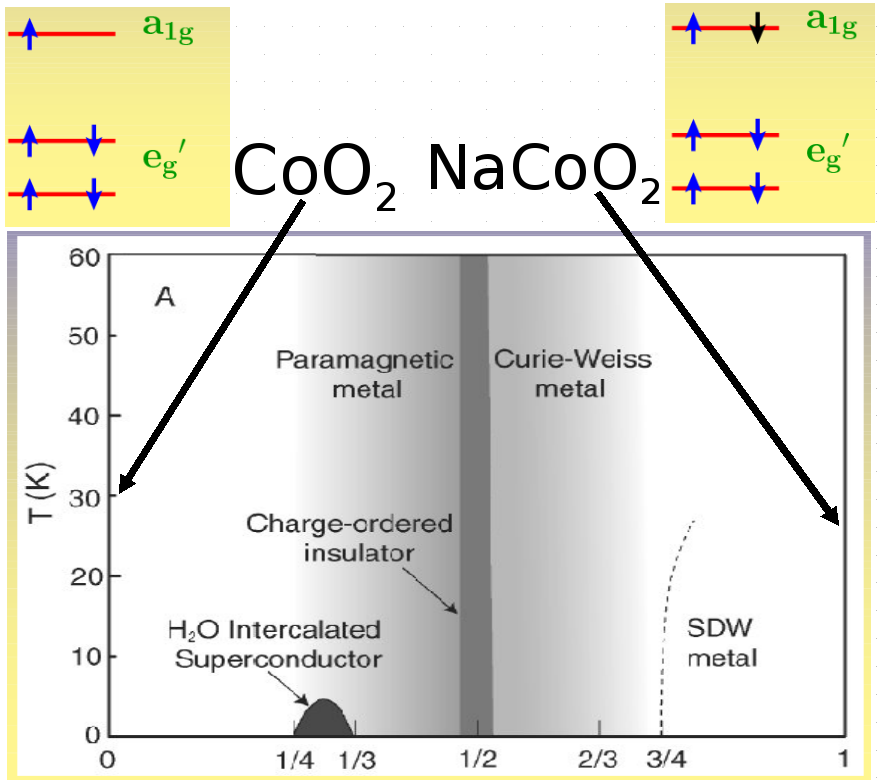}
\caption{ Phase diagram of CoO$_2$Na$_x$ compound. The Mott insulating
  side at $x=0$ has low thermopower, while the thermopower is greatly
  enhanced in the vicinity of the band insulator at $x=1$.  }.
\label{fig:7}
\end{figure}

Interest in thermoelectricity near the doping driven Mott transition
lead to theoretical and experimental investigations of
La$_{1-x}$Sr$_x$TiO$_3$ and CoO$_2$Na$_x$ for small values of the
concentration parameter $x$.  Both theory and experiment suggest that
the thermoelectric figure of merit is not very large in this regime.
On the other hand, the vicinity of the band insulator end,
La$_{1-x}$Sr$_x$TiO$_3$\cite{okuda} and CoO$_2$Na$_x$
(see Fig.~\ref{fig:7} for the phase diagram) were shown to have promising
thermoelectric performance. Should we conclude that Mottness is bad
for thermoelectricity?  Not necessarily, after all, clear signatures
of correlation were found in more realistic modeling of doped band
insulators once the impurity potentials of the dopant atoms were taken
into account \cite{dd}.  The impurity potential was found to restrict the
spatial regions available for the motion of the electricity and heat
carriers. In this restricted configuration space, the occupancy of the
electrons is close to integer and Mott physics is realized.

We have suggested that similar situation occurs in the electron gas
close to the metal insulator transition. Here, the long range
Coulomb interaction generates short range charge crystalline lattice
order. The occupancy of these lattice sites is close to integer
filling, suggesting that the character of the metal to insulator
transition is that of a Wigner Mott transition~\cite{vlad}. The
mechanism, spatial or orbital differentiation results in a restricted
low energy configuration, making Mott physics relevant. This mechanism
is quite general, and operates in other materials such as the
ruthenates~\cite{wang}. It could be called emergent Mottness or
super-Mottness, and contains similar physics to the orbital selective
Mott transition pheonomena. Hence (super)Mottness might be relevant
for high performance thermoelectricity after all!.  It would be useful
to reconsider the most recent advances in thermoelectric materials in
this light, and investigate the local magnetic susceptibility at the
impurity sites of the high performance thermoelectrics \cite{last,tags}.

\section{Outlook}

The outlook for material design in the field of thermoelectric is
quite promising. DMFT seems to capture qualitative trends in oxides of
practical interest, furthermore we have simple qualitative ideas, which
can be refined and tested with tools of ever increasing precision.  In
this context, the new thermoelectric modules to be developed in
conjunction with the new generation of LDA+DMFT codes, look very
appealing.  In conjunction with the renewed experimental efforts in
this field, the future looks very promising.

\begin{acknowledgement}
K.H. is supported by a grant of the ACS of the Petroleum Research
Fund. GK is supported by the DMR of the NSF.
\end{acknowledgement}

\input{referenc}

\end{document}

%% file: referenc.tex
%%%%%%%%%%%%%%%%%%%%%%%% referenc.tex %%%%%%%%%%%%%%%%%%%%%%%%%%%%%%